\documentclass[aps,prb,twocolumn,10pt,showpacs,amsmath,amssymb]{revtex4-1}
\usepackage{amsmath}
\usepackage{graphicx}
\usepackage{verbatim}
\usepackage{color}
\usepackage{subfigure}
\usepackage{hyperref}
\usepackage{soul}
\usepackage{amssymb}
\usepackage{cancel}
\newcommand{\A}{\mathbf{A}}

\newcommand{\rpi}[1]{\dot{\mathbf{r}}_{#1}}
\begin{document}
\title{Structural transitions in vertically and horizontally coupled parabolic channels of
 Wigner crystals}
\author{J. E. \surname{Galv\'an-Moya}}
\email[Email: ]{JesusEduardo.GalvanMoya@ua.ac.be}
\author{K. \surname{Nelissen}}
\email[Email: ]{Kwinten.Nelissen@ua.ac.be}
\author{F. M. \surname{Peeters}}
\email[Email: ]{Francois.Peeters@ua.ac.be}
\affiliation{Department of Physics, University of Antwerp, Groenenborgerlaan 171, B-2020,
 Antwerpen, Belgium}
%
\begin{abstract}

Structural phase transitions in two vertically or horizontally coupled channels of strongly
 interacting particles are investigated. The particles are free to move in the $x$-direction but
 are confined by a parabolic potential in the $y$-direction.  They interact with each other
 through a screened power-law potential ($r^{-n}e^{-r/\lambda}$).  In vertically coupled systems
 the channels are stacked above each other in the direction perpendicular to the $(x,y)$-plane,
 while in horizontally coupled systems both channels are aligned in the confinement direction.
 Using Monte Carlo (MC) simulations we obtain the ground state configurations and the structural
 transitions as a function of the linear particle density and the separation between the channels.
 At zero temperature the vertically coupled system exhibits a rich phase diagram with continuous
 and discontinuous transitions.  On the other hand the vertically coupled system exhibits only a
 very limited number of phase transitions due to its symmetry.  Further we calculated the normal
 modes for the Wigner crystals in both cases.  From MC simulations we found that in the case of
 vertically coupled systems the zigzag transition is only possible for low densities.  A
 Ginzburg-Landau theory for the zigzag transition is presented, which predicts correctly the
 behavior of this transition from which we interpret the structural phase transition of the Wigner
 crystal through the reduction of the Brillouin zone.
\end{abstract}
\pacs{ 05.20.-y, 64.60.F-, 81.30.-t }
\maketitle

\section{Introduction}

Self-organized systems are of fundamental importance in different areas of physics. can be seen
 in the accelerated progress in that area.  It all started in 1934 when Wigner\cite{045_wigner}
 surmised that, if considering an electron gas, where the electrons have no kinetic energy, as
 occurs in low density systems, then these electrons \textit{``would settle in configurations
 which correspond to the absolute minima of the potential energy.  These are closed-packed lattice
 configurations, with energies very near to that of the body-centered lattice.''}  This was the
 first prediction about self-organized systems of electrons and charged particles, which today is
 known as Wigner crystals.

At the present it is known that Wigner crystals form a body-centered cubic (BCC) lattice in
 three-dimensional\cite{008_hasse,005_cornelissens} (3D) space, a triangular lattice in
 two-dimensional\cite{004_schweigert,009_partoens} (2D) systems, while in one-dimensional (1D)
 systems, the energetically more favorable array is given by an evenly spaced lattice.  Wigner
 crystals have been the study object of several
 experiments\cite{020_itano,021_waki,022_ikegami,023_mortensen,033_dubin} and theoretical works
 \cite{006_piacente,014_fishman, 001_kwinten, 002_kwinten, 003_kwinten, 004_kwinten} in recent
 years.

Recently, the transition between 1D and 2D systems has been analyzed, known as
 quasi-one-dimensional (Q1D) systems.  Several studies\cite{003_piacente,006_piacente}
 have investigated different properties of the Wigner crystal in that regime.  When increasing the
 density, the linear chain structure undergoes a zigzag transition that occurs always through a
 continuous transition.  This zigzag transition has been observed in different
 experiments\cite{029_lutz,031_chan} and analyzed in different systems: colloidal
 systems\cite{027_saintjean,005_kwinten,006_kwinten,007_kwinten,008_kwinten}, vortices  in
 superconductor/ferromagnetic hybrids\cite{043_karapetrov}.  They have also been proposed as a
 possible implementation of a quantum information processor\cite{026_leibfried}.

The zigzag transition for a system of Coulomb particles subjected to a parabolic confinement
 potential\cite{011_delcampo} was analyzed through a Ginzburg-Landau theory.  Later this theory
 was applied to more complex particle interactions and generalized for a power-law confining
 potential\cite{041_galvan}.   In the latter work it was shown that this theory leads to a
 better representation of the phase transition behavior close to the critical point.

In the present work we analyze the phase transitions between two vertically and horizontally
 coupled Q1D channels as function of the inter-channel distance and linear particle density in
 each channel.  Further the zigzag transition of the vertically coupled channels is compared with
 the analytical results obtained from the Ginzburg-Landau approach.

The present paper is organized as follows. In Sect. \ref{model_system} the model system is
 formulated.  In Sect. \ref{numerical_model} the numerical results are presented with an analysis
 about the behavior of the system in the different structural phases.  In Sect. \ref{gl_model} the
 Ginzburg-Landau Lagrangian is derived and the theoretical results are discussed and compared with
 our numerical simulations.   Our conclusions are given in Sect. \ref{conclusions}.

\section{Model system}\label{model_system}

\begin{figure}[h!]
\begin{center}
\includegraphics[scale=0.50]{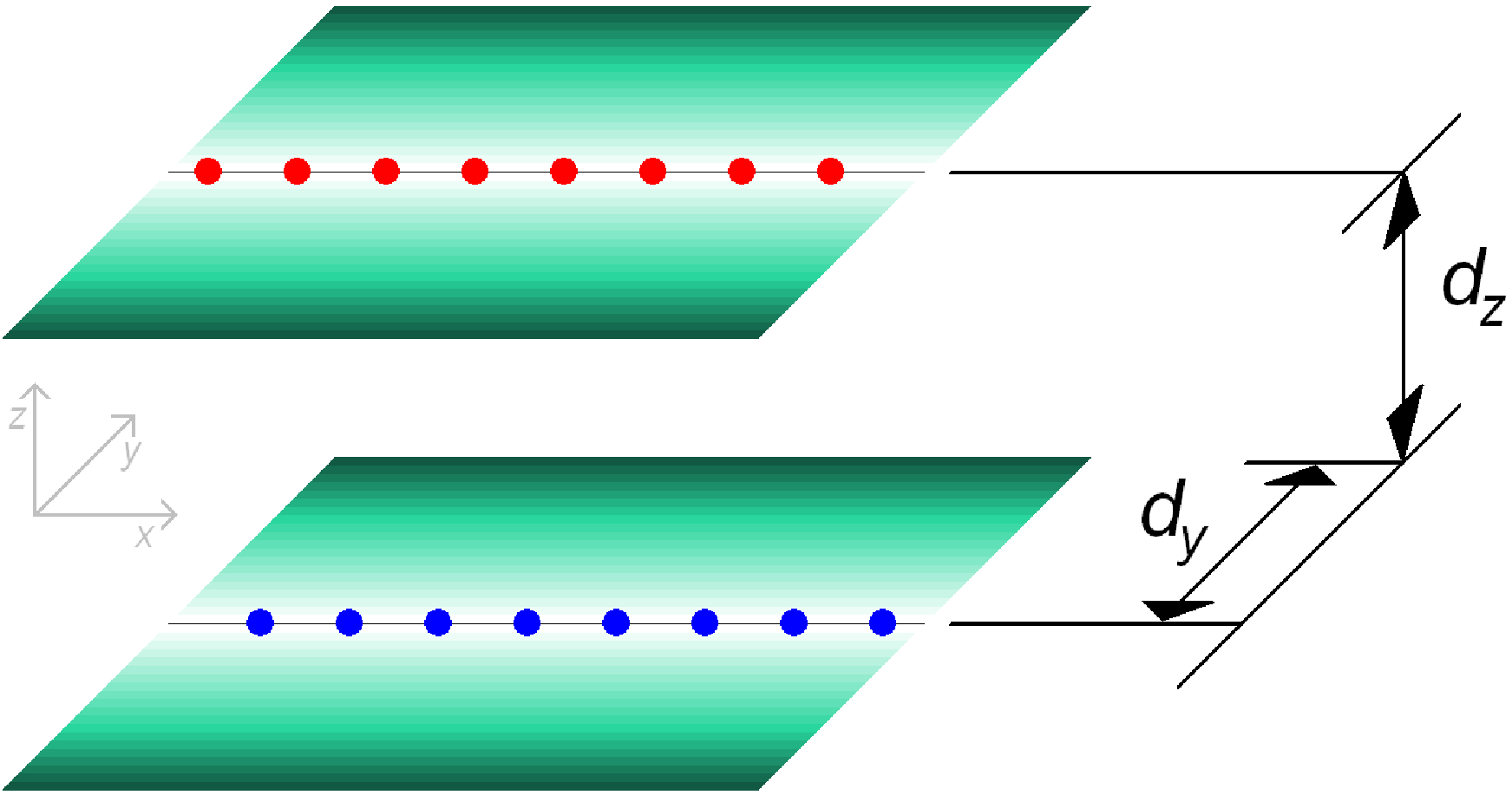}
\caption{\label{fig:model} (Color online) Schematic representation of the model system, where the filled circles
 represent the particles in each channel, $d_y$ and $d_z$ indicate the distance between the
 channels in the $y$- and $z$- axis respectively.  The darker colors in each plane represent the
 increases of the strength of the confinement potential in the channel.}
\end{center}
\end{figure}

In this work we consider a system consisting of two quasi-one-dimensional channels each with $N$
 identical particles with mass $m$ and charge $q$, which are restricted to move in the 
 $(x,y)$-plane. The system is subjected to an external magnetic field $\mathbf{B}$, which is applied in
 the direction perpendicular to the channels.  The particles are confined by a
 one-dimensional potential limiting their motion in the $y$-direction in each channel.  The
 channels are separated by a distance $d_y$ from each other in the $y$-direction and a distance
 $d_z$ in the $z$-direction as shown in Fig.~\ref{fig:model}.

The charged particles interact through a repulsive interaction potential.  The kinetic energy of
 the system is given by:
\begin{equation}\label{total_energy}
    T = \sum_{i=1}^{N}{\frac{\mathbf{p}_{1i}^2}{2m}} +
	\sum_{i=1}^{N}{\frac{\mathbf{p}_{2i}^2}{2m}},
\end{equation}
\noindent where $\mathbf{p}_{ki}=m\mathbf{v}_{ki}+q\mathbf{A}_{ki}$ represents the canonical
 momentum of particle $i$ in channel $k$, which is moving with velocity $\mathbf{v}_{ki}$.~$\mathbf{A}_{ki}$ is the vector potential related to the magnetic field through the relation
 $\mathbf{B}=\nabla\times\mathbf{A}_{ki}$.  The total energy of the system is given by:
\begin{eqnarray}
  E & = &  \frac{m}{2}\sum_{i=1}^{N} (\rpi{1i}^2+\rpi{2i}^2)
	    + q\sum_{i=1}^{N} (\rpi{1i}\cdot\A_{1i} + \rpi{2i}\cdot\A_{2i}) \nonumber\\
	& & + \frac{q}{2m}\sum_{i=1}^{N} (\A_{1i}^2 + \A_{2i}^2)
	    + V_{conf} + V_{int},
\end{eqnarray}
where the confinement and the interaction potential are, respectively, given by:
\begin{subequations}\label{Pot_Energy_original}
    \begin{equation}
        V_{conf} =
	    \frac{1}{2}m\upsilon_{t}^2\sum_{i=1}^{N} \left(y_{1i}-\frac{d_y}{2}\right)^{2} +
	    \frac{1}{2}m\upsilon_{t}^2\sum_{i=1}^{N} \left(y_{2i}+\frac{d_y}{2}\right)^{2},
    \end{equation}
    \begin{eqnarray}
        V_{int} & = & \sum_{i=1}^{N}\sum_{j>i}^{N} V_{pair}(r_{1i,1j})
		    + \sum_{i=1}^{N}\sum_{j>i}^{N} V_{pair}(r_{2i,2j}) \nonumber\\
		& & + \sum_{i=1}^{N}\sum_{j=1}^{N} V_{pair}(r_{1i,2j}).
    \end{eqnarray}
\end{subequations}

Here, $V_{pair}(r)$ represents the inter-particle potential which is taken as a screened power-law
 potential, which will allow for the simulations of both long- and short-range interactions, as
 follows:
\begin{equation}\label{screened_power_pot}
    V_{pair}(r) = \frac{q^2}{\epsilon R} \frac{R^{n} e^{-r/\lambda}}{r^{n}}.
\end{equation}

In general $r_{Ai,Bj}=|\mathbf r_{Ai}-\mathbf r_{Bj}|$ represents the relative position between
 the $i$-th particle in channel $A$ and the $j$-th particle in channel $B$.  The exponent $n$ is
 an integer and $\epsilon$ is the dielectric constant of the medium where the particles are moving
 in.  In the above, $R$ is an arbitrary length parameter which we introduced to guarantee the correct
 units.  The magnetic field is taken to be constant and is applied in the direction perpendicular to the
 plane formed by the channels, through the vector potential defined by
 $\A_{ki}=-By_{ki}\mathbf{e_x}$.  The total energy can be written in dimensionless form as
 follows:
\begin{eqnarray}\label{Total_Energy_dimensionless}
    E &=&   \sum_{i=1}^{N} (\rpi{1i}^2+\rpi{2i}^2)
	  - B\sum_{i=1}^{N} (\dot{x}_{1i}y_{1i} + \dot{x}_{2i}y_{2i}) \nonumber \\ & &
	  + \left(\upsilon^2 + \frac{B^2}{4}\right)\sum_{i=1}^{N}(y_{1i}^2 + y_{2i}^2) \nonumber\\ & &
	  + d_y\upsilon^2 \sum_{i=1}^{N}\left(y_{1i} - y_{2i} - \frac{d_y}{2}\right) \nonumber\\ & &
          + \sum_{i=1}^{N}\sum_{j>i}^{N} \frac{e^{-\kappa r_{1i,1j}}}{ r_{1i,1j}^{n}}
	  + \sum_{i=1}^{N}\sum_{j>i}^{N} \frac{e^{-\kappa r_{2i,2j}}}{ r_{2i,2j}^{n}}\nonumber\\ & &
	  + \sum_{i=1}^{N}\sum_{j=1}^{N} \frac{e^{-\kappa r_{1i,2j}}}{ r_{1i,2j}^{n}},
\end{eqnarray}
where the dimensionless frequency is given by $\upsilon=\upsilon_{t}/\omega_0$, while $\omega_0$
 measures the strength of the confinement potential and $t_0=1/\omega_0$ is the unit
 of time.~The unit of the energy is given by: $E_0=\left(m\omega_0^2/2\right)^{n/(n+2)}
 \left(q^2/\epsilon\right)^{2/(n+2)}R^{2(n-1)/(n+2)}$, the distances are scaled with
 $r_0=\left(2q^2/m\omega_0^2\epsilon\right)^{1/(n+2)} R^{(n-1)/(n+2)}$, and the strengths of the
 magnetic field and the vector potential are measured in units of $B_0 = 2m\omega_0/q$ and
 $A_0=B_0r_0$, respectively.  Additionally, the dimensionless parameter $\kappa=r_0/\lambda$
 represents the screening parameter of the potential.  In order to describe better the behavior
 of the transitions in the system, we define a dimensionless linear density $\eta$ as the
 number of particles per unit of length along the unconfined direction in each channel.  This
 density is chosen to be equal in both channels.

\section{Numerical Model}\label{numerical_model}

In order to understand the behavior of coupled systems, we consider two different cases.  Case A
 corresponds to two vertically coupled channels where the distance between the channels in the
 $y$-direction is taken zero.  In this case the channels are arranged above each other.  Case B
 consists of two horizontally coupled channels with $d_z=0$, which are aligned parallel to each
 other separated by a distance $d_y$.  In order to find the ground state configuration for each
 case we have performed Monte Carlo simulations, complemented with the Newton optimization
 technique\cite{004_schweigert,002_schweigert}.  Using those methods we construct the phase
 diagram of the ground state configuration of the system at zero temperature as a function of the
 separation between the two channels and the linear
 density.

\subsection{Case A: Vertical Coupling}

\begin{figure} [ttt]
\begin{center}
\includegraphics[scale=0.70]{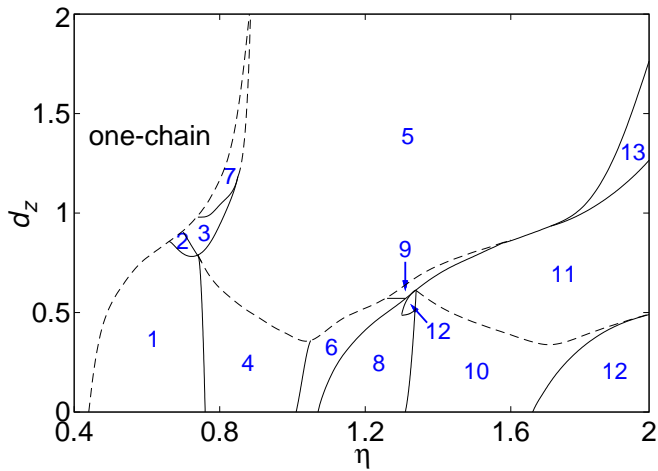}
\caption{\label{fig:z_phase_diagram} Phase diagram of vertically coupled channels as a function of
 the linear particle density and the distance between the channels in the $z$-direction.  Solid
 and dashed lines represent first and second order transitions, respectively.  The different phases
 are discussed in the text and shown in Fig.~\ref{fig:z_confs}.}
\end{center}
\end{figure}

The phase diagram of vertically coupled channels is shown in Fig.~\ref{fig:z_phase_diagram} for
 an interaction potential with parameters $n=1$ and $\kappa=1$ and fixed value of the frequency 
 $\upsilon=1$.  The phases are represented by the number in each delimited region, the solid lines
 represent first order transitions and the dashed lines second order transitions.

\begin{figure} [ttt]
\begin{center}
\includegraphics[scale=0.25]{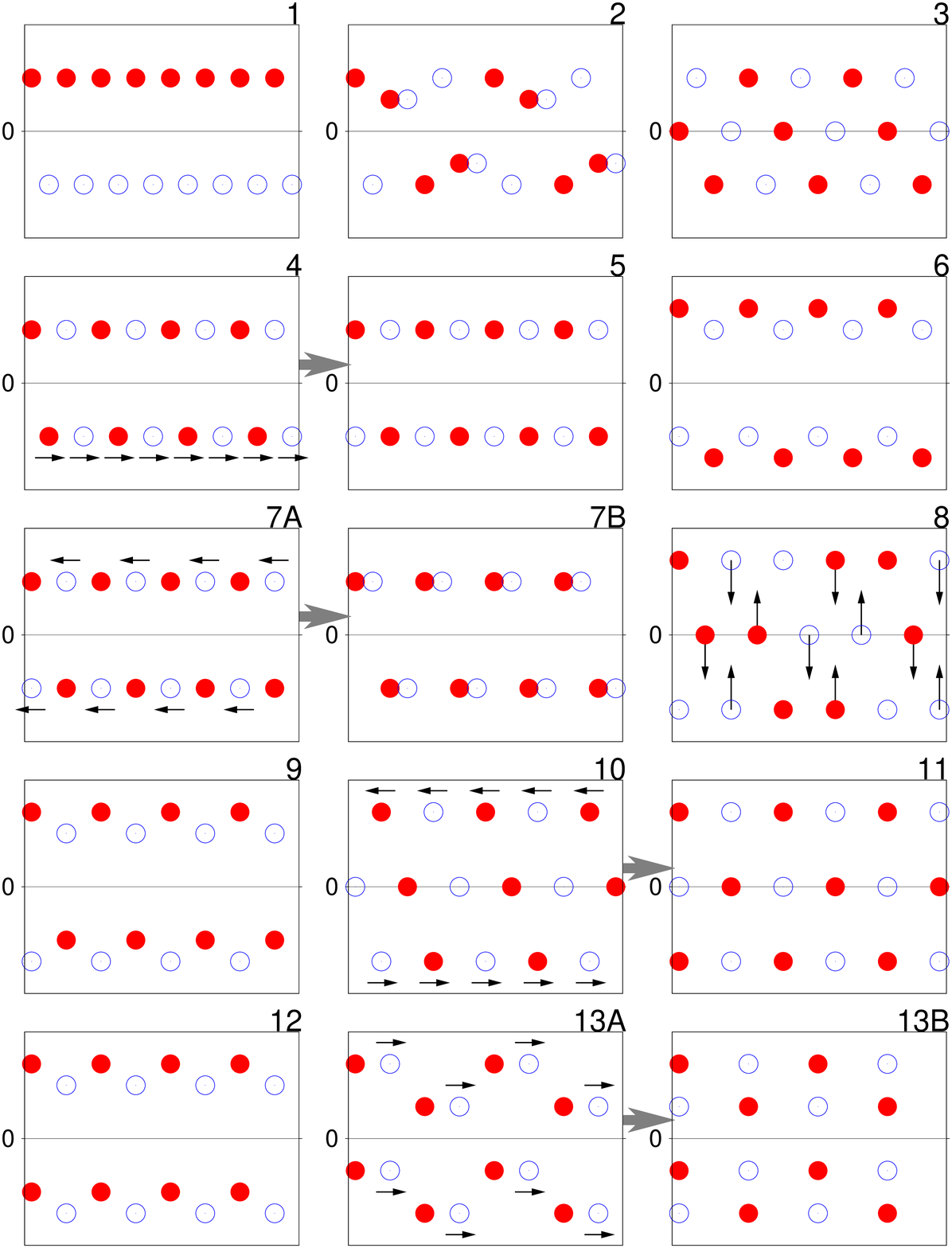}
\caption{\label{fig:z_confs} (Color online) Configurations of the different crystalline phases of
 the particles as enumerated in the phase diagram presented in Fig. \ref{fig:z_phase_diagram}.
 The red solid (blue open) symbols are the particles in the top (bottom) channel.}
\end{center}
\end{figure}

From this phase diagram we see that the zigzag transition occurs only between the one-chain
 configuration and phase $1$ where particles of each channel are organized in a single chain,
 this transition is continuous and of second order.
 The different phases are shown in Fig.~\ref{fig:z_confs}.  One can see that the activation 
 point, defined as the point where the one-chain arrangement is no longer the only possible ground
 state configuration for any value of $d_z$, is given by $\eta_{act}=0.447302$ for $d_z=0$.  This
 value of the linear density represents exactly half the value of the critical density ($\eta_c$)
 for the zigzag transition in the case of a single Q1D channel as was found in Refs.
 \onlinecite{006_piacente,041_galvan}.  Notice that for increasing $d_z$, the critical density
 increases as well, which means a change in the total dimensionality of the system. Additionally,
 one can observe that for $d_z>0.5$, $\eta_c$ increases more quickly than for $d_z<0.5$.  This
 implies that the system tends to behave as two decoupled channels, until the point where the
 zigzag transition disappears.

The different ground state configurations identified in Fig. \ref{fig:z_phase_diagram} are shown
 in Fig. \ref{fig:z_confs}.  In this figure, the red filled circles represent the particles in
 channel A and the blue open circles the particles in channel B.  The gray arrows between two
 configurations indicate that the transition between them occurs continuously, following the
 process described through the small black arrows inside each phase configuration.  The small
 black arrows above each particle represent the displacement direction when the separation between
 the channels is increased.  It is interesting to observe that the phases $7$, $8$ and $13$
 are composed of  several configurations whose transition occurs through a continuous phase
 transition indicated by the small black arrows plotted inside the pictures.

\begin{figure} [ttt]
\begin{center}
\includegraphics[scale=0.25]{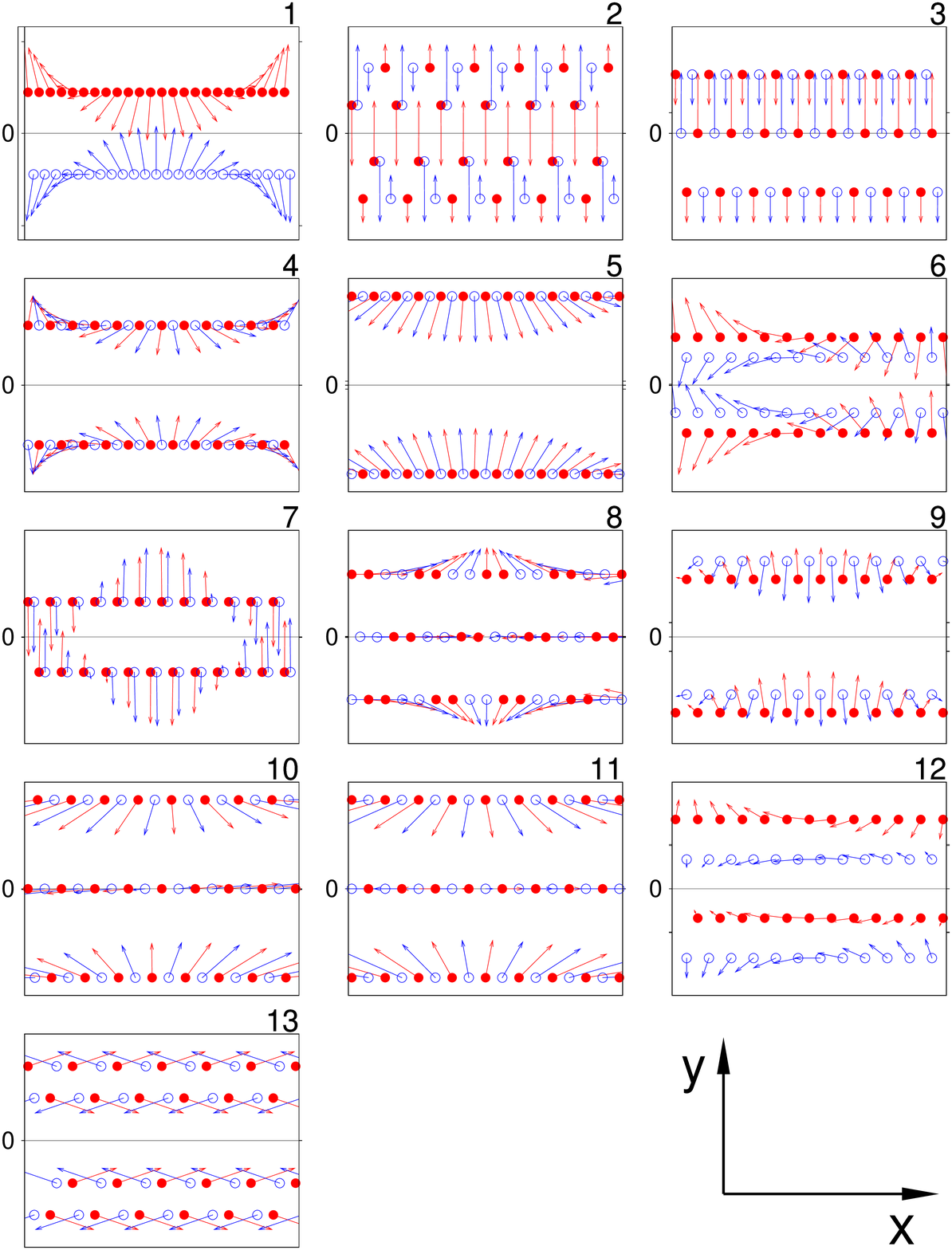}
\caption{\label{fig:z_modes} (Color online) Eigenvectors of the first non-zero frequency of each
 phase for vertically coupled systems.}
\end{center}
\end{figure}

The eigenvectors of the first non-zero eigenfrequency for each ground state configuration is shown
 in Fig. \ref{fig:z_modes}.  The direction of the eigenvector is of particular interest
 because it contains information about the onset of melting in that crystal, when temperature is
 increased.

\begin{figure} [h!]
\begin{center}
\includegraphics[scale=0.70]{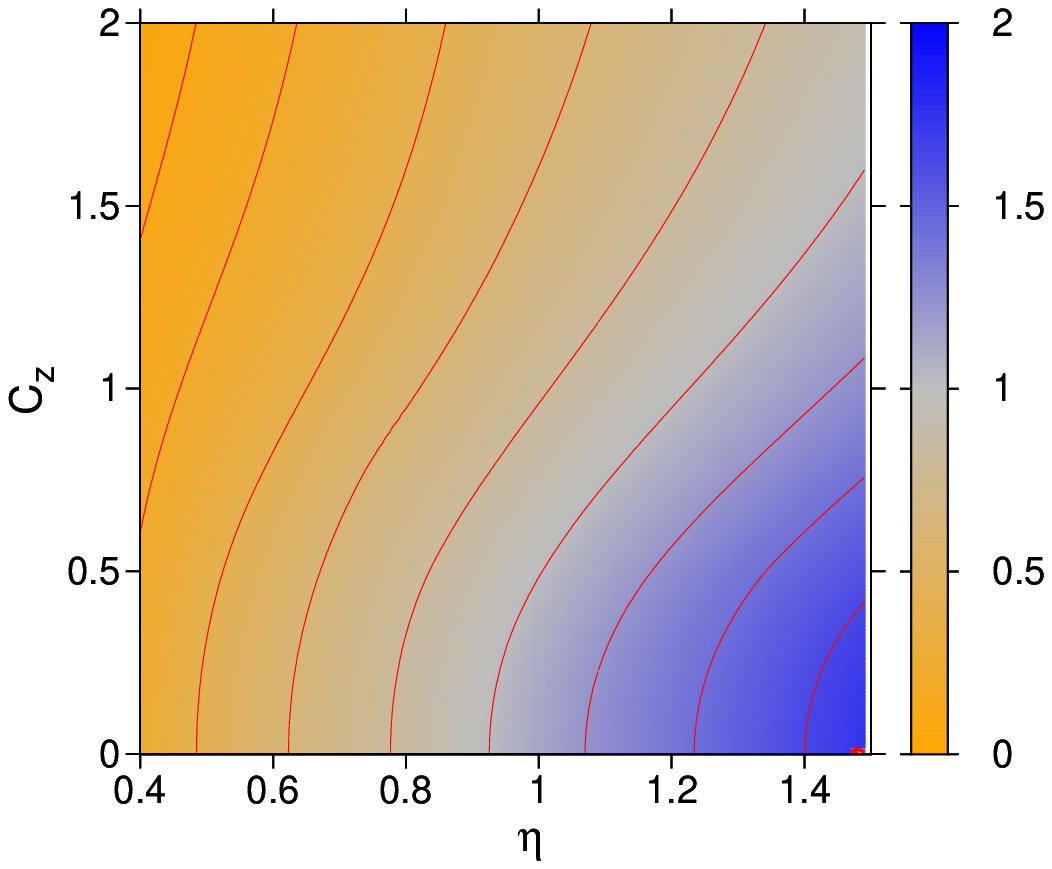}
\caption{\label{fig:z_energy3D} (Color online) Contour plots of the energy per particle as a function of the
 linear density $\eta$ in each channel and the vertical separation $d_z$ between the channels.
 The red lines represent the iso-energy lines of the system.}
\end{center}
\end{figure}

\begin{figure} [hb!]
\begin{center}
\includegraphics[scale=0.60]{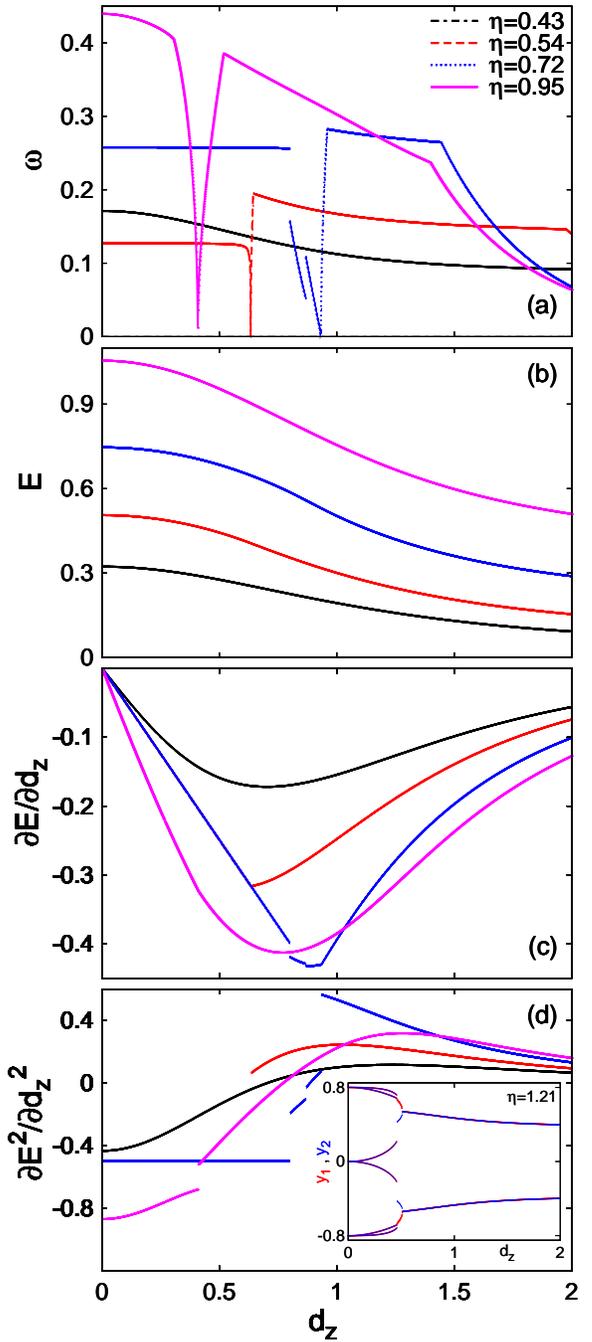}
\caption{\label{fig:z_freqen} (Color online) (a) First non-zero eigenfrequency; (b) Energy per particle;
 (c) First derivative of the energy; (d) Second derivative of the energy.  In all figures we have
 plotted as a function of the vertical separation between the channels. The colors showed at the
 top represent the different densities. The inset in (d) shows the distance of the chains
 from the $x$-axis of the channels as a function of the vertical separation for $\eta=1.21$.}
\end{center}
\end{figure}

In order to know the evolution of the system it is interesting to show the behavior of the
 total energy as a function of the control parameter.  This can be seen in
 Fig.~\ref{fig:z_energy3D}, where the contour plots of the energy per particle are plotted as a
 function of the linear density and the vertical separation between channels.  The color scale
 indicates the magnitude of the energy and the lines show the iso-energy lines on the plot.
 Notice that by increasing the density, the iso-energy lines become increasingly curved showing
 the decoupling of the channels, due to the increase of the interaction between particles in the
 same channel, while the energy contribution from the interaction between channels becomes
 significantly smaller with increasing $d_z$.

\begin{figure} [ttt]
\begin{center}
\includegraphics[scale=0.70]{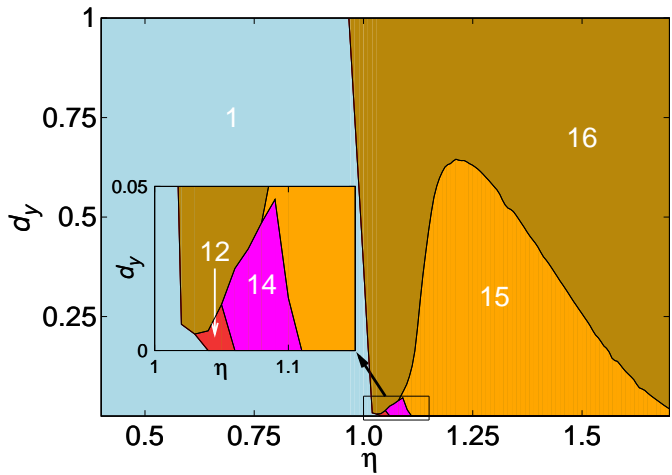}
\caption{\label{fig:y_phase_diagram} (Color online) Phase diagram of the system as a function of the linear
 density ($\eta$) and the distance between the channels in $y$-direction ($d_y$). All phase
 transitions are first order transitions.}
\end{center}
\end{figure}

\begin{figure} [b]
\begin{center}
\includegraphics[scale=0.70]{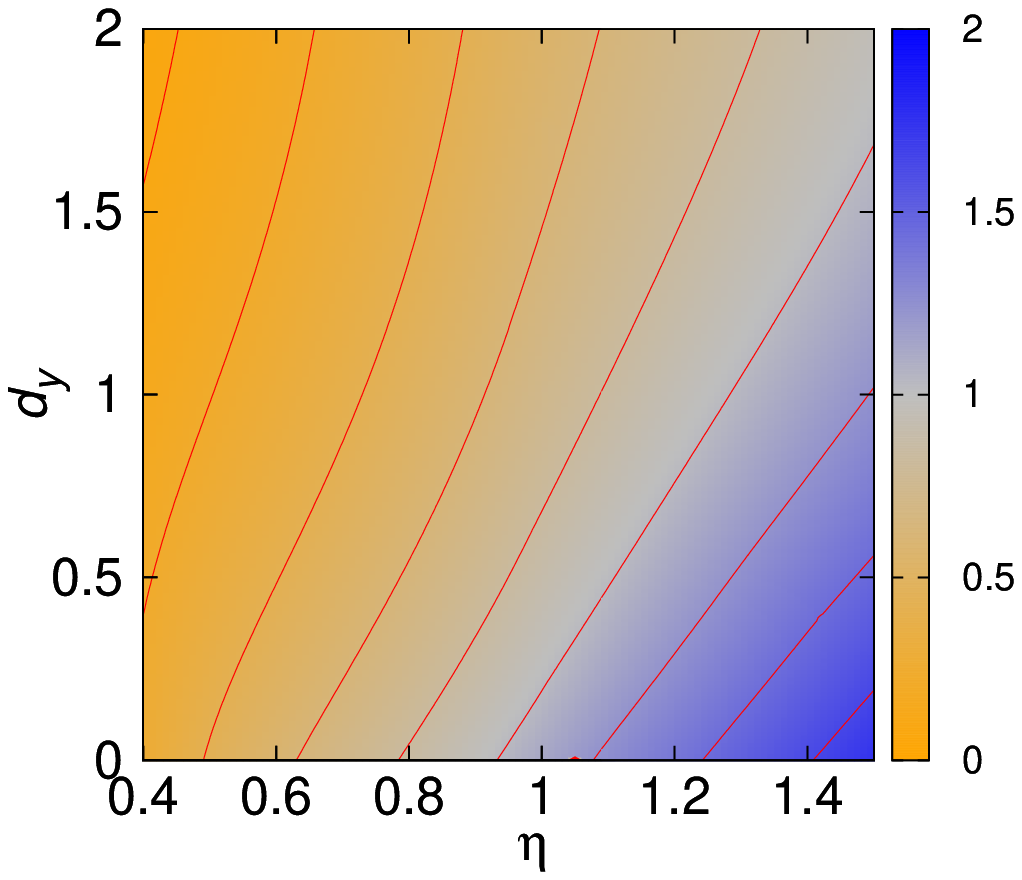}
\caption{\label{fig:y_energy3D} (Color online) Contour plots of the energy per particle as a function of the
 linear density $\eta$ in each channel and the horizontal separation $d_y$ between channels.  The
 red lines represent the iso-energy lines of the system.}
\end{center}
\end{figure}

\begin{figure} [hbp!]
\begin{center}
\includegraphics[scale=0.60]{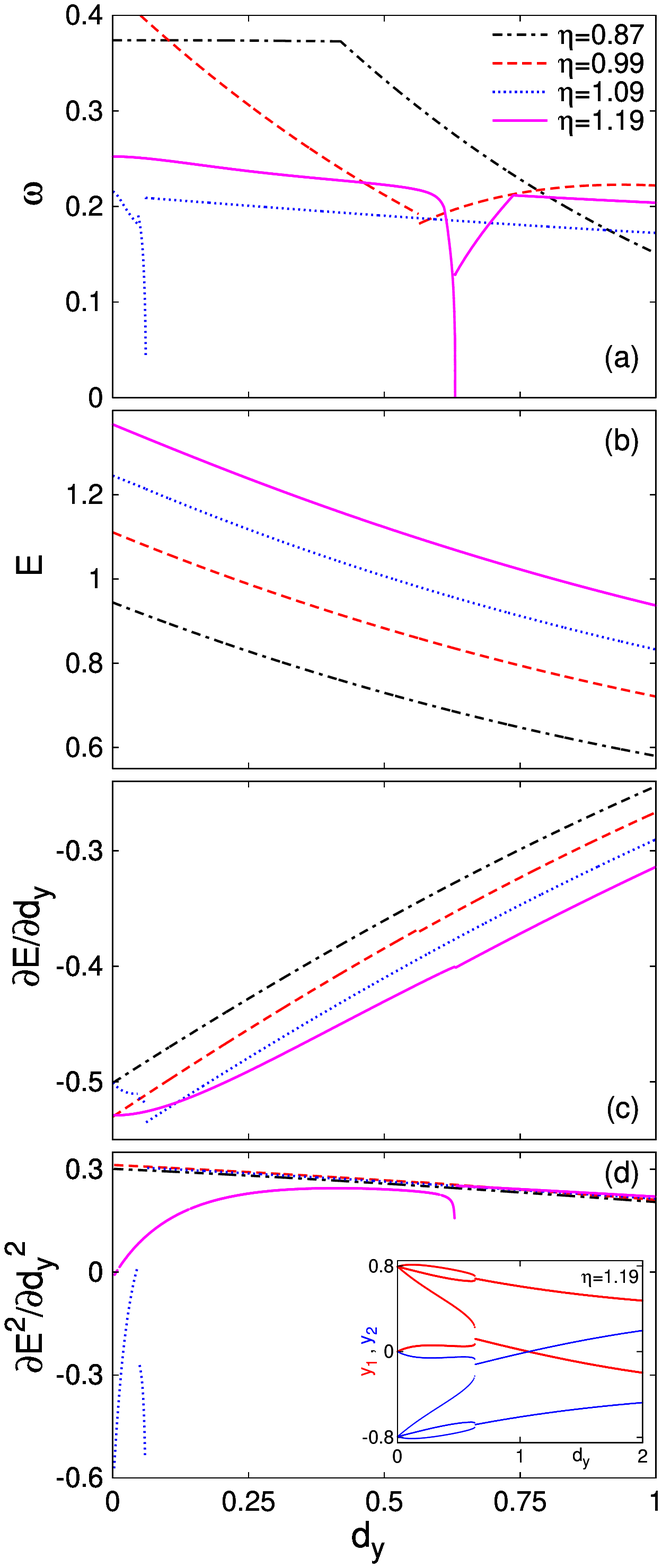}
\caption{\label{fig:y_freqen} (Color online) (a) First non-zero eigenfrequency; (b) Energy per particle;
 (c) First derivative of the energy; (d) Second derivative of the energy.  All figures are plotted as a function of the horizontal
 distance between the channels.  The colors showed at the top represent the different densities.
 The inset in (d) shows the distance of the chains from the center of each channel as a function
 of the horizontal separation for $\eta=1.19$.}
\end{center}
\end{figure}

With the aim to explain the order of the transitions, we plot in Fig.~\ref{fig:z_freqen} the first
 non-zero eigenfrequency for different densities.  We also plot the energy per particle and its
 two first derivatives with respect to the vertical separation.  In all cases the figures show the
 behavior of these quantities as a function of the separation between channels, and the value of
 the density taken for each curve is printed at the top of the figure.~From the smooth and
 monotonic behavior of the energy in Fig.~\ref{fig:z_freqen}(b) and the jumps in its derivatives (see
 Figs.~\ref{fig:z_freqen}(c,d)), we can deduce the order of each phase transition, which following
 the Ehrenfest classification \cite{B03_Uzunov} are given by the order of the lowest derivative of
 the energy which exhibits a discontinuity.  From Fig.~\ref{fig:z_freqen}(a) we note that the
 transitions between phase~1 and phase~2 (shown for $\eta=0.72$ around $d_z=0.798$) and between
phase 2 and phase 3 (shown for $\eta=0.72$ around $d_z=0.866$) are of first order which can be
 seen by the jumps in the first derivative of the energy at the transition points. The transitions
 of second order are characterized by the softening of an eigenmode, i.e. $\omega=0$.  The inset
 in Fig.~\ref{fig:z_freqen}(d) shows the $y$-position of each chain for $\eta=1.21$, where one can see the continuous transition from
 $3$ to $6$ chains, the first order transition between $6$ and $4$ chains and the second and
 continuous transitions between $4$ and $2$ chains.  The red and blue lines represent the positions
 of the particles in channel A and channel B.

\subsection{Case B: Horizontal Coupling}

In this part we analyze horizontally coupled channels ($d_z=0$) for the same parameters as in case
 A.  The phase diagram in this case is shown in Fig.~\ref{fig:y_phase_diagram} as a function of
 the linear density $\eta$ and the vertical separation between the channels $d_y$.

From this phase diagram we notice that the zigzag transition does not occur in horizontally
 coupled channels. The one-chain configuration only exists when $d_y=0$ and for $\eta<\eta_{act}$.
 Additionally the small number of phases is a result of the weaker interaction between channels.

In Fig.~\ref{fig:y_energy3D} we show a contour plot of the energy per particle and iso-energy
 lines as a function of the linear  density and the horizontal separation between the channels.
Notice the smooth and monotonic behavior of the iso-energy lines, which align parallel to each
other with increasing linear density.~When increasing $d_y$, the interaction between the
channels decreases quickly due to the separation between channels which is directly related with the
 confinement potential as can be seen in Eq.~(\ref{Total_Energy_dimensionless}), where $d_y$
 contributes to the total energy of the system introducing a linear term of the relative position
 between particles in the $y$-direction.  This contribution leads to the disappearance of the
 harmonicity of the system.

In Fig.~\ref{fig:y_freqen}(a) we show the first non-zero frequency as a function of the horizontal
 separation for different densities, which shows clearly the phase transition points.  The
 monotonic and continuous behavior of the energy (Fig.~\ref{fig:y_freqen}(b)) and the jumps in the
 first derivative of the energy, as shown in Fig.~\ref{fig:y_freqen}(c), indicates, that all phase
 transitions are of first order.  In those transition points we also can see that the first
 non-zero eigenfrequency has a jump, which shows the abrupt change in the vibration mode of
 the system in the phase transitions. As an illustration about the first order transitions in the
 system, we show in the inset of Fig.~\ref{fig:y_freqen}(d) the $y$-distance of the chains from the
 center of each channel, where one can see the transitions between $3$, $8$ and $4$ chains for
 $\eta=1.19$. In this figure the red and blue lines represent the particles in channel A and
 channel B, respectively.

\begin{figure*} [htpb!]
\begin{center}
\includegraphics[scale=0.25]{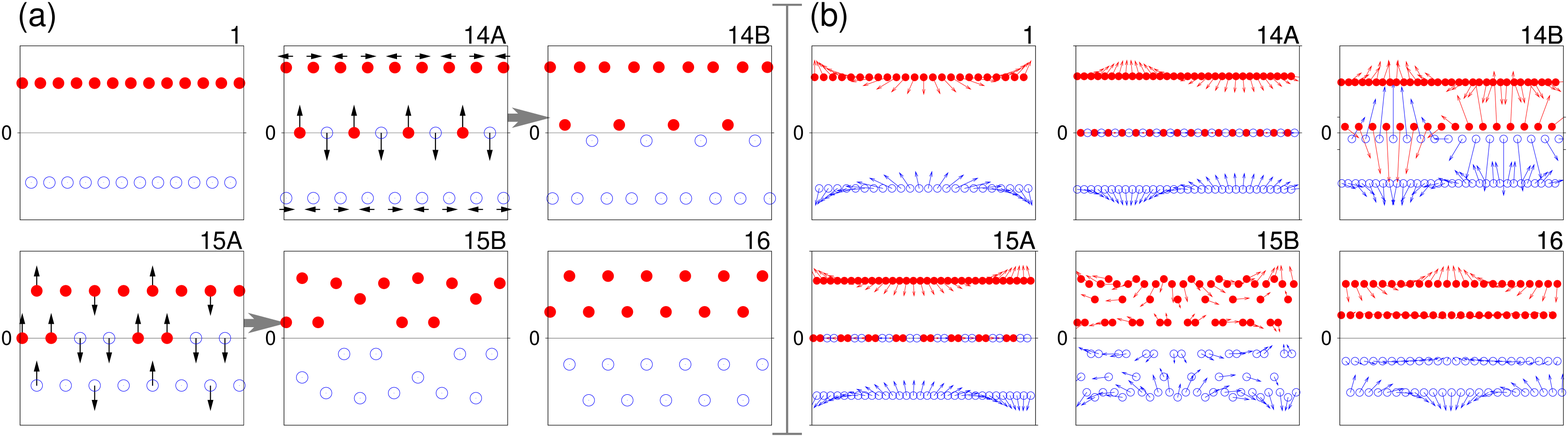}
\caption{\label{fig:y_confmodes} (Color online) (a) Configurations of the crystalline
 organization of particles enumerated in the phase diagram presented in
 Fig.~\ref{fig:y_phase_diagram}.  (b) Eigenvector of the first non-zero frequency of each phase
 configuration.}
\end{center}
\end{figure*}

The different configurations of the phases shown in the phase diagram of the horizontally
 coupled system, are plotted in Fig.~\ref{fig:y_confmodes}(a), where the gray arrows and the small
 black arrows have the same meaning as previously.  Notice that phases $14$ and $15$ consist of
 several phases.  The transition between those phases occurs through continuous transitions.  In
 Fig.~\ref{fig:y_confmodes}(b) the normal mode of the first non-zero frequency is plotted for each
 configuration.

\section{Ginzburg-Landau Model for linear-zigzag transition}\label{gl_model}

The results of previous section indicates that the transition between the one-chain arrangement and
 the zigzag configuration occurs in the low density region and only in the case of a vertically
 coupled system.  To analyze this transition (i.e. $d_y=0$), we start by considering the system in
 the situation that the one-chain configuration is stable in each channel and close to the zigzag
 transition point.  The equilibrium positions of all particles are given by
 $\mathbf{r}_{ki}^{lin}=(x_{ki}^{lin},0,(k-1)d_z)$ with $x_{ki}^{lin}=[2i+(2-k)]/2\eta$, where $k$
 represents the channel number (i.e. $1$ or $2$) and $i$ the particle number in that channel.
 Further we consider small oscillations around the equilibrium positions in the $y$-direction.
 The position of each particle becomes thus
 $\mathbf{r}_{ki}=\mathbf{r}_{ki}^{lin} + \Delta\mathbf{r}_{ki}$, where the small displacements
 are given by $\Delta\mathbf{r}_{ki}=(0,y_{ki},0)$.  Now, ensuring that the particles are
 oscillating around these ground state positions, and assuming that these oscillations are much
 smaller than the distance between the two nearest particles, we expand $r_{1i,2j}$ into a power
 series.

As found numerically, this transition is continuous and the configuration after the transition
 point consists of a zigzag configuration (phase $1$ in Fig. \ref{fig:z_phase_diagram}), but with
 all particles forming a single off-center chain in each channel (i.e. $y_{1i}=c$ and $y_{2i}=-c$,
 with $c$ as a real number).  From that configuration we can observe that there is a competition
 between the interaction potentials $V_{11}$, $V_{22}$ and $V_{12}$ which are defined as the last
 three terms in Eq. (\ref{Total_Energy_dimensionless}) respectively.  The subindexes represent the
 interacting channels.  The first two interaction potentials ($V_{11}$, $V_{22}$) do not
 contribute to the solution in this case, since all particles in the same channel are aligned
 parallel to the $x$-axis.  Thus, the interaction between the particles in the different channels,
 as well as the confinement potential, are responsible for this transition.

Following Ref. \onlinecite{041_galvan}, we obtain the Lagrangian density of the system in the
 continuum limit by considering only the interaction potential $V_{12}$:
\begin{eqnarray}\label{lagrangian}
 \mathcal{L} = \frac{1}{2} &\Big[&  h^2(n,\tilde\kappa,c_z)(\partial_x \psi(x))^2
				+ \delta_{\upsilon}(n,\tilde\kappa,c_z)\psi^2(x) \nonumber\\ & &
				+ \mathcal{A}(n,\tilde\kappa,c_z)\psi^4(x) \Big],
\end{eqnarray}
where $\tilde\kappa = \kappa/\eta$, $c_z=\eta d_z$,
 $\delta_{\upsilon}(n,\tilde\kappa,c_z)=2(\upsilon^2+B^2/4)-\varpi^2(n,\tilde\kappa,c_z)$ and
 the order parameter $\psi(x)$ represents the separation between channels (i.e. $\psi=2c$).  The
 coefficients are given by:
\begin{widetext}
\begin{eqnarray}
 h^2(n,\tilde\kappa,c_z) &=& \eta^n\Bigg[ 
		n              \sum_{j=0}^{\infty}\frac{e^{-\tilde\kappa R_j}}{R_j^{n  }}\left(1-2\sin^2\left(\frac{R_j\tilde k_0}{2}\right)\right)
                + \tilde\kappa \sum_{j=0}^{\infty}\frac{e^{-\tilde\kappa R_j}}{R_j^{n-1}}\left(1-2\sin^2\left(\frac{R_j\tilde k_0}{2}\right)\right)
				 \Bigg] \\
 \varpi^2(n,\tilde\kappa,c_z) &=& 4\eta^{n+2}\Bigg[ 
		n              \sum_{j=0}^{\infty}\frac{e^{-\tilde\kappa R_j}}{R_j^{n+2}}\sin^2\left(\frac{R_j\tilde k_0}{2}\right)
                + \tilde\kappa \sum_{j=0}^{\infty}\frac{e^{-\tilde\kappa R_j}}{R_j^{n+1}}\sin^2\left(\frac{R_j\tilde k_0}{2}\right)
				 \Bigg] \\
 \mathcal{A}(n,\tilde\kappa,c_z) &=& 2n\eta^{n+4}\Bigg[ 
		\frac{n+2}{2}  \sum_{j=0}^{\infty}\frac{e^{-\tilde\kappa R_j}}{R_j^{n+4}}\sin^4\left(\frac{R_j\tilde k_0}{2}\right)
                + \tilde\kappa \sum_{j=0}^{\infty}\frac{e^{-\tilde\kappa R_j}}{R_j^{n+3}}\sin^4\left(\frac{R_j\tilde k_0}{2}\right)
				 \Bigg],
\end{eqnarray}
\end{widetext}
where $R_j=\sqrt{(j+1/2)^2+c_z^2}$ and $\tilde k_0=k_0/\eta$ represent the value of the wavevector
 in the stability point.

\subsection{Stability Point}

The stability point is calculated numerically, as the location of the minimum of the second order
 term of the total energy, and the value of the wavevector is called $k_0$.  Then the stability
 point is found from the condition $\min \delta_{\upsilon}(n,\tilde\kappa,c_z)$.
 Notice that the value of $\tilde k_0$ depends on the distance between channels.  For $c_z=0$ we
 can find analytically that $\tilde k_0 = 2\pi$, which implies that the particle density ($\eta$)
 of the system in each channel  will be reduced to half of the density required to find the
 stability point in one channel as found previously\cite{006_piacente,041_galvan}.  Next, from the
 stability condition we find for $\upsilon=1$ that the activation point (minimum value of
 density to find the zigzag transition) is $\eta_{c(min)}=0.447302$.
\begin{figure} [hb!]
\begin{center}
\includegraphics[scale=0.70]{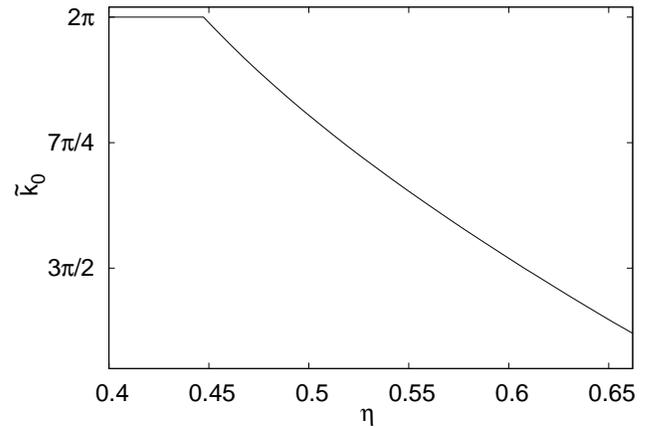}
\caption{\label{fig:k0_density} Wavevector corresponding to the second lowest eigenvalue as a
 function of the particle density in each channel. System parameters: $\upsilon=1$, $n=1$, and
 $\kappa=1$}.
\end{center}
\end{figure}

We show in Fig.~\ref{fig:k0_density} the behavior of $\tilde k_0$ as a function of the particle
 density in each channel.  Notice that for densities lower than $\eta_{c(min)}$, which represents
 the region of the one-chain configuration, the value is $\tilde k_0=2\pi$.  This value indicates
 that the first Brillouin zone is twice as large as in the one channel case.  This value is
 related to $\eta$, because the total linear density of the system is $2\eta$ when $c_z=0$ and it
 decreases by increasing $\eta$.  For lower densities ($\eta<\eta_{c(min)}$) the linear
 arrangement is preserved due to the weaker interaction potential between the particles as
 compared to the confinement potential and this for any separation between the channels.  However,
 when $\eta>\eta_{c(min)}$, the zigzag configuration is found to result in a lower interaction
 energy between the particles in the different channels. This behavior can be reduced by
 increasing $c_z$ and therefore reducing by the total density of the system.  In that situation
 $\tilde k_0$ is of major importance, because while the separation increases and the total
 density decreases the first Brillouin zone becomes smaller reducing the dimensionality of the
 crystal in each channel.  This behavior is shown in Fig.~\ref{fig:k0_density} where one can see
 the monotonic decrease of $\tilde k_0$ with $\eta$, when the linear density exceeds the critical
 density.

In order to find the ground state configuration, we minimize Eq.~(\ref{lagrangian}) and we obtain
 the equation of motion of the system.  Then, after minimization, the solution is reduced to the
 well-known analytical expression
 $\psi=\sqrt{-\delta_{\upsilon}(n,\tilde\kappa,c_z)/2\mathcal{A}(n,\tilde\kappa,c_z)}$ which is
 valid when the density is lower than the critical density, as was found in
 Ref.~\onlinecite{041_galvan} for parabolic confinement.

\begin{figure} [ttt]
\begin{center}
\includegraphics[scale=0.70]{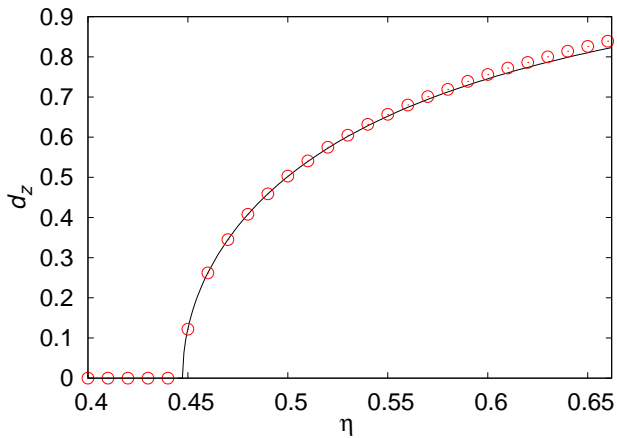}
\caption{\label{fig:dz_density} (Color online) Critical value of distance between channels as a
 function of the particle density in each channel.  The solid line is the theoretical result and
 the open circles are the results from our MC simulations. We took the parameters $\upsilon=1$,
 $n=1$, and $\kappa=1$}
\end{center}
\end{figure}

Fig.~\ref{fig:dz_density} represents the critical separation between channels as a function of
 the particle density.  This critical value corresponds to the separation $c_z$ where the zigzag
 transition takes place.  This critical separation is calculated as the lower value of $c_z$ at
 which $\psi=0$ for a given $\eta$.  For points below the curve in Fig.~\ref{fig:dz_density}
 the zigzag arrangement is the ground state configuration.  The open red circles are the results
 for the zigzag transition points calculated from our MC simulations.  The plotted region
 corresponds to the region where the zigzag transition is allowed, as was analyzed in previous
 section.  One can see that there is perfect agreement between our theoretical results and the
 results from the MC simulations.

\begin{figure} [h!]
\begin{center}
\includegraphics[scale=0.70]{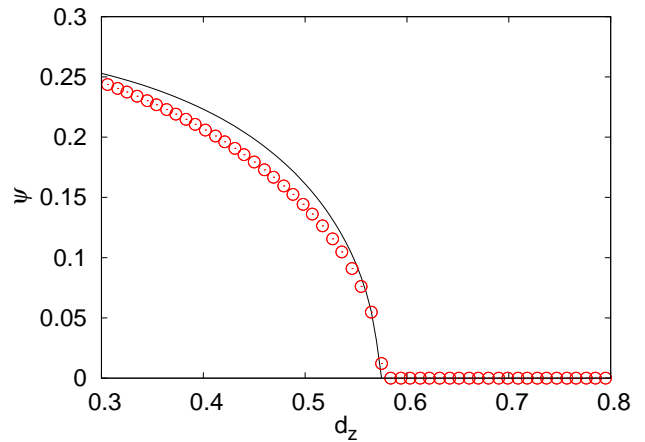}
\caption{\label{fig:psi_density} (Color online) Distance between channels as a function of the
 vertical separation between the channels, where the linear density in each channel is $\eta=0.52$.  We took the
 parameters $\upsilon=1$, $n=1$, $\kappa=1$.}
\end{center}
\end{figure}

In Fig.~\ref{fig:psi_density} we show the distance between chains in the confinement direction
 as a function of the separation between the channels $c_z$ for a linear density $\eta=0.52$.
 The open circles show the results from our MC simulations and the solid line represents the value
 of the order parameter $\psi$ calculated from the present theory.  The good agreement indicates 
 that the Ginzburg-Landau theory is able to describe correctly the behavior of the system close to
 the zigzag transition point.

\section{Conclusions}\label{conclusions}

We studied two coupled Q1D channels, consisting of a system of interacting charged particles
 confined by a parabolic trap in each channel, where the linear particle density in both channels
 is given by $\eta$.  The ground state configuration at zero temperature was analyzed.

The structural transitions between phases were studied as a function of the linear density and the
 separation between the channels.  We found a very rich phase diagram in case of vertically
 coupled channels, with first and second order transitions.  The horizontally coupled system, on
 the other hand, exhibits a very restricted number of phases and all transitions are of first
 order.  The latter can be traced back to the linear term in the relative distance between
 particles in the different channels, which appears in the energy expression.  This linear term
 results in a rapid decrease of the interaction between channels.

Our simulations show that the zigzag transition occurs only in case of vertically coupled
 channels, which represent another effect of the linear term in the energy.  In order to
 understand the behavior of this zigzag transition, we derived a Ginzburg-Landau equation and
 determined the behavior of the system close to the zigzag transition point.  We analyzed the
 order parameter and its dependence on the linear density and the separation between channels.
 From theory we found that, for vertically coupled channels, an increase of the  density produces
 a reduction of the first Brillouin zone of the Wigner crystal after the activation point
 $\eta_{act}$.~ Within the density range where the zigzag transition is allowed, the vertical
 separation could be used as a tunable parameter to modulate the transition point.  This result
 was found from theory and there is a perfect agreement with our numerical results, which shows
 that the presented theory is sufficient to understand the behavior and the nature of the zigzag
 transition in Q1D coupled channels.

\section{Acknowledgments}\label{thanks}
This work was supported by the Flemish Science Foundation (FWO-Vl).

\end{document}